# Realization of Stable Ferromagnetic Order in Topological Insulator: Codoping Enhanced Magnetism in 4$f$ Transition Metal Doped Bi$_2$Se$_3$

Bei Deng[1], Yiou Zhang[1], S. B. Zhang[2], Yayu Wang[3], Ke He[3] and Junyi Zhu[1,*]

[1]*Department of physics, the Chinese University of Hong Kong, Hong Kong SAR*

[2]*Department of Physics, Applied Physics, and Astronomy, Rensselaer Polytechnic Institute, Troy, New York 12180, USA*

[3]*Department of Physics, Tsinghua University, Beijing 100084, People's Republic of China*

The realization of long range and insulating ferromagnetic states in topological insulator (TI) has been a pressing issue since its discovery. Only recently, such state was achieved in Cr-doped Bi$_{2-x}$Sb$_x$Te$_3$, leading to the discovery of quantum anomalous Hall effect (QAHE). However, the effect is only observed at extremely low temperatures mainly due to the limited magnetism. To fully understand the mechanism of the ferromagnetic ordering whereby improving the ferromagnetism, we investigated 4$f$ transition metal-doped Bi$_2$Se$_3$, using density-functional-theory approaches. We found that Eu and Sm prefer the Bi substitutional sites with large magnetic moments to ensure stable long-range ferromagnetic states. Additionally, codoping can be a novel strategy to preserve the insulating property of the host material as well as improving the incorporation of magnetic dopants. Our findings thus offer the critical step in facilitating the realization of QAHE in TI systems.

The discovery of the long-sought quantum anomalous Hall state (QAHS) in topological insulators (TIs) has attracted tremendous research interests in recent years[1-4]. Long range, stable, and carrier-free ferromagnetism is critical to realize QAHS in TI materials[5-8]. Typically, Bi$_2$Se$_3$ and similar compounds are promising candidates of the second generation TIs for such application[9-15]. However, the magnetic doping mechanism of such materials is still not well understood[7,16], and a stable ferromagnetism remains to be achieved[17].

Generally, the difficulty in achieving a stable ferromagnetic ordering in Bi$_2$Se$_3$ rises from the following reasons: i) the traditional magnetic impurities, such as 3$d$ metals, normally exhibit low solubilities[17-19] on the substitutional site due to their large size-mismatch with the host; ii) The interaction between $d$-$d$ bands of the impurities tend to induce the formation of clusters that may destroy the long-range ordering[20-22]; and iii) the unfavorable super-exchange mechanism often leads to a transformation from ferromagnetism to non-ferromagnetism or anti-ferromagnetism. For instance, it was reported that Cr-doped Bi$_2$Se$_3$ can be both ferromagnetic (FM)[21] and anti-ferromagnetic (AFM)[23], implying a competition between the FM and AFM couplings among the Cr impurities. The super-exchange mechanism was later verified by an $ab$-initio study[17], in spite that the ground state of Cr-doped Bi$_2$Se$_3$ is FM. As a result, the total energy of the FM state is only about 10 meV lower than that of an AFM state, resulting in a low Curie temperature[1].

Besides the 3$d$ transition metals (TMs), 4$f$ TMs could be promising candidates to realize stable long-range ferromagnetism in Bi$_2$Se$_3$. Because they possess atomic radii comparable to the host Bi atoms, the size-mismatch will not be a severe problem. For example, the ionic radius[24] of 4$f$ TM Eu$^{3+}$ is 0.95 Å, which is closer to that of Bi$^{3+}$ (1.03 Å), compared with that of Cr$^{3+}$ (0.62 Å). Also, the $f$ bands are more localized and less interactive than the $d$ bands, naturally suppressing the formation of impurity clusters.

Another advantage of the 4$f$ elements is attributed to their large atomic mass, which gives stronger spin-orbit coupling (SOC) strength than 3$d$ elements. From the viewpoint of van Vleck mechanism, this is better for the FM order. Furthermore, as the $f$ bands have more electronic orbitals than the $d$ bands have, they could possess even higher magnetic moment and hence a stronger FM coupling. Until now, however, a systematic investigation is still lacking.

In this study, to address the above questions, we investigated the dopabilities and magnetism of 4$f$ elements in Bi$_2$Se$_3$ by $ab$ $initio$ calculations, based on density functional theory (DFT) [see computational details in Supplementary Information (SI)]. We found that the formation energy of 4$f$ TM impurities on the Bi sites is lower than that on the interstitial sites. Several 4$f$ substitutional impurities were found to possess large magnetic moments and strong FM coupling. Among these impurities, Eu and Sm show very large magnetic moment of 6.33 and 5.06 $\mu$B, respectively, and both with strong FM character. The calculated energy difference between the FM state and AFM state is about 3-4 times larger than that of Cr$_{Bi}$, as reported in previous works[17]. Besides Eu and Sm, Tb and Dy are also promising candidates for their large magnetic moment, moderately-stable FM state, and low formation energy. Moreover, to realize long range and insulating FM states, we proposed a novel method, i.e., codoping electron donors with the impurities such as Eu. This way, the Fermi level can be tuned, while the formation energy of the impurities can be lowered. These findings can have important implications on the realization of the quantum anomalous Hall effect (QAHE).

We first consider TMs at the interstitial and Se sites, as these impurities, TM$_i$ and TM$_{Se}$, often cause large changes in the electronic structure, whereby leading to a deterioration of the magnetic ordering, and hence should be avoided. Our calculation reveals that the 4$f$ impurities are unlikely to occupy these detrimental sites. Consider for example, the TM$_i$s. There are two types of them in the Bi$_2$Se$_3$ lattice, namely, the octahedral and tetrahedral sites, each with 6 different configurations. We have examined all of them and found that the most stable one is on the octahedral site between two QLs. Yet, it's significantly more difficult to form TM$_i$ than to form TM$_{Bi}$, as the formation energy of TM$_i$ is at least 1.08 eV higher than that of TM$_{Bi}$ at the most favorable Bi-rich conditions. This difference is noticeably larger than the reported 0.29 eV for the 3$d$ TM impurities[17], implying that the 4$f$ TMs are very unlikely to occupy any of the interstitial sites. This is mainly due to the relatively large sizes of the 4$f$ TMs that better match Bi than the 3$d$ TMs do.

Next, we consider eight selected TM$_{Bi}$ impurities, which are Nd, Sm, Eu, Gd, Tb, Dy, Ho, and Er. The calculated formation energies are shown in Fig. 1 as a function of the Se chemical potential. Except for Eu$_{Bi}$, Sm$_{Bi}$ and Ho$_{Bi}$, the calculated formation energies are always negative, regardless the growth conditions. This is in line with experimental observation that a number of the 4$f$ elements can form continuous alloy with Bi$_2$Se$_3$ and/or Bi$_2$Te$_3$[25-28]. It is noted that Eu and Gd have a significantly large difference in the formation energy (> 1.3eV), even though they are adjacent neighbors in the Periodic Table. As this difference cannot be understood solely by the atomic-size difference, we have studied the effect of atomic chemical potentials on the formation of these impurities (see Fig. S1 in SI). We found that to avoid the formation of secondary phases, in particular, TM selenides, the maximum Eu chemical potential at equilibrium is -4.11 eV, which is 0.43 eV lower than that of Gd (-3.68 eV). The reason for the lower chemical potential of Eu lies in its exceptionally stable secondary phase EuSe, where Eu has a valence +2, leaving behind with a very stable half-occupied $f$ band, according to the Hund's rule. The lower chemical potential for Eu here is consistent with the larger impurity formation energy. The remaining formation energy difference between Eu and Gd may be attributed to their different electronic structures – while Gd$_{Bi}$ has occupied $f$ bands deeply buried inside the valence band, Eu$_{Bi}$ has its partially occupied bands near the Fermi level, and hence Eu$_{Bi}$ cannot be nearly as stable as Gd$_{Bi}$.

To understand their magnetic behavior, we calculated the magnetic moments of all 4$f$ TMs and TABLE I lists the results. These results suggest that the crystal filed splitting of those 4$f$ impurities is generally weak in Bi$_2$Se$_3$, associated with the corresponding exchange splitting (see SI for detailed discussions on crystal field splitting and exchange splitting). We note that among all the candidates, Gd$_{Bi}$ and Eu$_{Bi}$ have the largest magnetic moments, being 7$\mu$B and 6.33$\mu$B, respectively, in the direction perpendicular to Bi$_2$Se$_3$ QLs. These values are greater than the maximum magnetic moment of 5$\mu$B ever possible for 3$d$ TM impurities[17].

A stable long-range magnetic order depends not only on the local magnetic moment, but also on the coupling of the magnetic moments in the same direction. To estimate the magnetic coupling in a solid, one usually places two TM impurities in a supercell, with a set of different distances, and then calculate the energy difference between the FM and AFM states (defined here as $\Delta E_{FM-AFM}/2$). By using this approach, we found that the strongest coupling between TM impurities typically happens at the second nearest neighbor distance within the same QL (whose energy difference, when significant, is given in TABLE I), while the coupling of impurities between QLs can be noticeably smaller. The results show that the FM coupling is favorable in Sm- and Eu-doped Bi$_2$Se$_3$ and moderately favorable in Tb- and Dy-doped Bi$_2$Se$_3$, while the Er- and Gd-doped Bi$_2$Se$_3$ favors a weak AFM coupling. Note that massive Dirac fermion has been observed experimentally in Dy-doped Bi$_2$Te$_3$ up to room temperature[28], implying the existence of a long-range FM order. Nd in TABLE I represents a different case in which the nearest neighbor FM coupling (21 meV) is the strongest, but the second nearest neighbor coupling is only 1 meV.

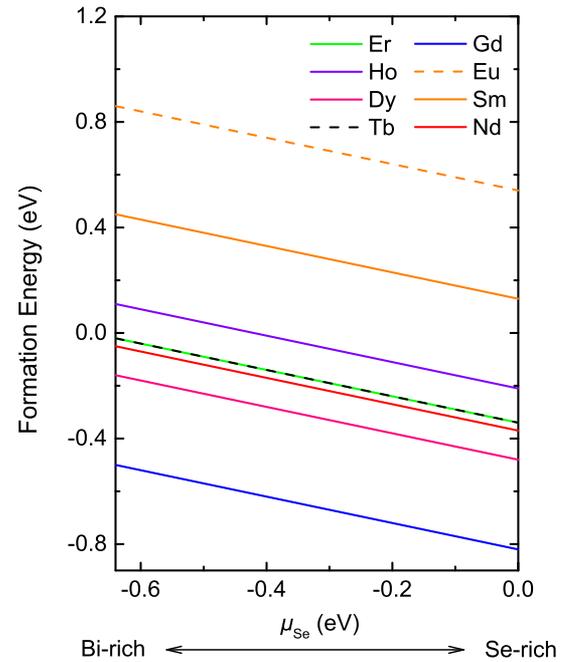

FIG 1. Calculated formation energy of eight TM$_{Bi}$ impurities, as a function of the chemical potential of Se.

To further our understanding, we compare the band structure and projected density of states (PDOS), shown in Fig. 2(a)-(f), for Eu and Gd. The band gap for defect-free Bi$_2$Se$_3$ is 0.30 eV. This value is reduced to 30 and 79 meV, respectively, for Eu- and Gd-doped systems (see SI for detailed discussion on the reliability of the obtained band gap). The gap narrowing phenomenon may be understood as a combined result of the SOC and $p$-$f$ couplings near the valence band maximum (VBM). It is known that the 4$f$ states are usually more localized than the $3d$ states. This is typified by the highly-localized impurity states inside the band gap for Eu, as can be seen in Figs. 2 (b) and (e). In contrast, for Gd, the impurity bands are much deeper, about -3.5 to -4.5 eV below the Fermi level, as revealed by Fig. 2 (f), due to the extra $f$ electron of Gd. While according to the Hund's rule, having 7 electrons half-occupying the $f$ shell makes the Gd impurity very stable, the lower $f$ bands also make the $p$-$f$ coupling less significant than the Eu-doped case, so the band gap of the former is also noticeably larger than the latter.

TABLE I. Magnetic moments and magnetic coupling strengths ($\Delta E_{FM-AFM}/2$) of eight 4$f$ TM impurities.

| Dopant | La | Ce | Pr | Nd | Sm | Eu | Gd | Tb | Dy | Ho | Er | Tm | Yb | Lu |
|---|---|---|---|---|---|---|---|---|---|---|---|---|---|---|
| Magnetic Moment ($\mu$B) | 0 | 0.71 | 1.95 | 3 | 5.06 | 6.33 | 7 | 6.01 | 4.99 | 3.98 | 2.92 | 1.90 | 0.76 | 0 |
| $\Delta E_{FM-AFM}/2$ (meV) | – | – | – | -1 | -44 | -31 | 2 | -14 | -14 | -10 | 3 | – | – | – |

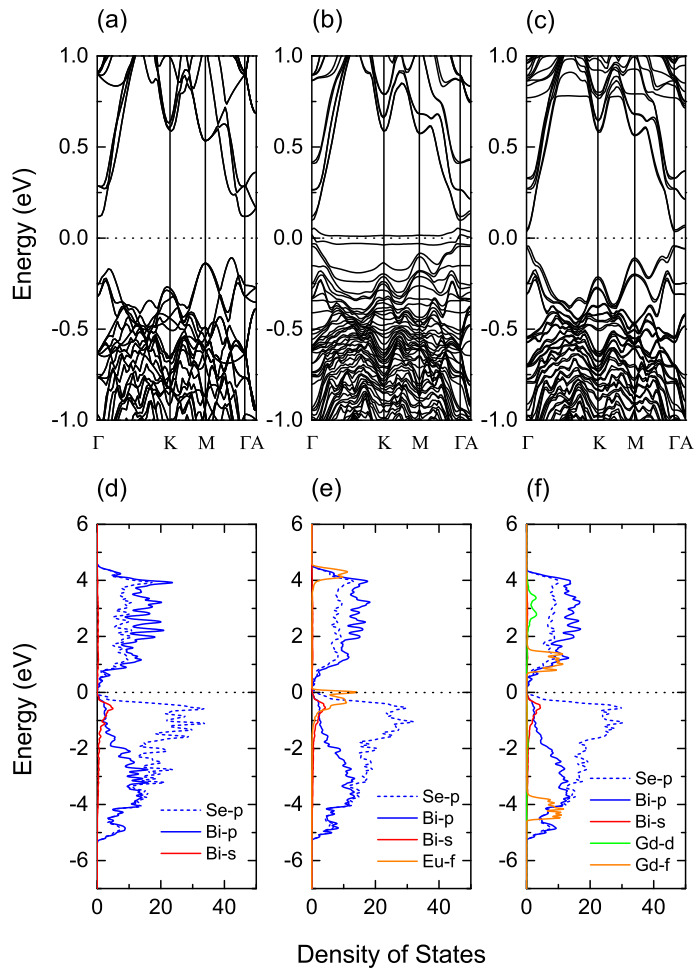

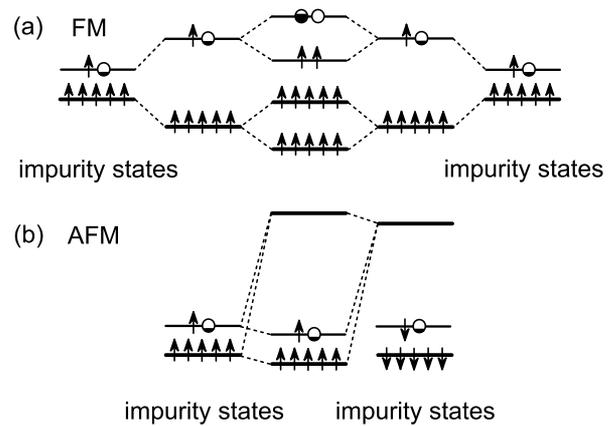

FIG. 2. Band structure for (a) pure, (b) Eu-doped, and (c) Gd-doped $Bi_2Se_3$ and PDOS for (d) pure, (e) Eu-doped, and (f) Gd-doped $Bi_2Se_3$. Fermi level is set to zero, as indicated by the dash lines. SOC has been included in the calculation.

FIG. 3. Schematic illustration for the (a) ferromagnetic and (b) anti-ferromagnetic couplings in Eu-doped $Bi_2Se_3$.

The band structures in Fig. 2 provide us with a qualitative picture for the formation of the FM and AFM states in Eu- and Gd-doped $Bi_2Se_3$, which could also be extended to other systems. The strong FM coupling in Eu-doped system is suggested to originate from the exchange coupling between Eu impurities via the intervening Se atoms. Because the partially occupied Eu state is above the VBM, the double exchange mechanism (i.e., the coupling between the $d$-$d$ or $f$-$f$ bands of the impurities) dominates in this configuration. In the FM states, the coupling between the impurity bands lowers the fully occupied majority-spin states [see Fig. 3(a), the outmost panels], which in turn couple to the $4p$ states (of the Se atoms) at the VBM. This coupling also raises the partially-occupied majority-spin impurity states near the Fermi level relative to the fully occupied majority-spin states [see Fig. 3(a), the second outmost panels]. Then, coupling between the partially-occupied impurity states, across the impurities [see Fig. 3(a), the middle panel] lowers the total energy of the system. In the AFM states, in contrast, coupling between the low-lying occupied majority-spin states [i.e., those in Fig. 3(b), the outmost panels] and the high-lying unoccupied minority-spin states in the conduction band takes place [see Fig. 3(b), the middle panel]. While such a coupling can also lower the total energy of the system, the exchange splitting between the majority and minority spins is quite large for $4f$ TM impurities, which counteracts and leads to a smaller energy gain. Therefore, the FM state is the ground state in Eu-doped $Bi_2Se_3$. This, however, does not happen in Gd-doped system, because the spin-up impurity states are deep inside the valence band. An FM coupling on fully occupied states cannot gain the system much energy. Instead, the AFM coupling will be more favorable. This rationale is supported by the calculated weak AFM state for Gd-doped $Bi_2Se_3$ in TABLE I, which is also supported by experiment[25].

Moreover, we noticed that Sm also favors a strong FM state, with a large magnetic moment of 5.06 $\mu B$, and a lower impurity formation energy than Eu. Also, the FM coupling strength in Sm-doped $Bi_2Se_3$ is 44 meV, even stronger than Eu. From a previous calculation by Chen *et al.*[29], in conjunction with the calculated PDOS in Fig. 2 (e) for $Eu_{Bi}$, we see that $Sm_{Bi}$ has an electronic structure similar to $Eu_{Bi}$, but with one less $f$ electron. The coupling between the lower-lying fully-occupied impurity bands and the higher-lying partially-occupied impurity bands with one less electron generally should lead to a larger energy gain. These results provide interpretation for the recent experiment [29], where stable FM state in 5% Sm-doped $Bi_2Se_3$ samples was observed at 52K.

To realize QAHE in a TI, a successful magnetic doping requires that the as-doped system remains to be insulating, the dopants-induced magnetic moments align in the direction perpendicular to the quintuple layer, and that the magnetic dopants have a sufficient concentration to achieve long-range FM order. Therefore, to realize QAHE in Eu- or Sm-doped $Bi_2Se_3$, which is $p$-type, as revealed by Fig. 2 (e) in this paper and Fig. 5(c) in Ref. 29, it is essential to be able to raise the Fermi level into the band gap. One possible strategy is to incorporate with the TM the right amount of $n$-type dopants, such as group VII anions or group IV cations. As such a codoping can also change the occupation of the $f$ bands, the magnetic moments of the dopant, or the coupling strength between the FM and AFM states, its overall effect needs be addressed.

We calculated the magnetic moment and the magnetic coupling strength for Eu codoped with the donors, which range from F to I for group VII elements, and Ge, Sn, and Pb for group IV elements. We found that $Eu_{Bi}$, in general, has relatively larger magnetic moments but lower ferromagnetic coupling strength, when codoped with donors (see TABLE S1 in SI). It suggests that the FM coupling strength can be sensitive to charge compensation by the codopants, as a result of changing the occupancy of the Eu impurity bands. However, isolated donors cannot create magnetic moment in the absence of Eu, thus the enhanced magnetic moment cannot be interpreted in terms of the spin-spin interaction between the donors. Instead, we note that there is a charge transfer from the *p* states of the donors to the *s*, as well as the majority-spin *f* states of Eu. A larger occupation of the $Eu_{Bi}$ spin-up states results in the larger magnetic moment. On the down side, however, this charge transfer also leads to a larger occupation of the partially-occupied impurity states, whereby raising the total energy of the magnetically-coupled system and in turn lowering its coupling strength. This explains why the FM coupling strength of Eu is always lowered by codoping with donors. Nonetheless, the donor-compensated $Eu_{Bi}$ impurity still has a reasonably strong FM coupling strength above 20 meV, suggesting that the FM ground state of the Eu can be preserved, irrespective of the exact concentration of the co-dopant. This insensitivity is due to the fact that the exchange coupling is mediated by spin-polarized holes *localized* at the impurity states, subject to bounded magnetic polarons (BMPs)[30-32]. To verify this, we performed further calculations on a donor-compensated Sm impurity. The calculated FM coupling strength is also on the order of 20 meV, in line with the Eu results. In our view, however, $Sm_{Bi}$ may not be as promising as $Eu_{Bi}$ for realizing QAHE in $Bi_2Se_3$, because Sm has one less *f* electron than Eu, so two electron donors are needed to fully compensate one Sm.

Finally, we need to address the issue of $Bi_2Se_3$ doping by Eu. The incorporation of Eu is expected to be more difficult than Cr, given that the calculated minimum formation energy for Eu is 0.54 eV, which is significantly higher than that of Cr (about -0.25 eV)[17]. This, as we discussed earlier, can be understood in terms of the formation of competitive secondary phases and the energy position of the *f* bands. With the help of codoping, the +2 valence of Eu may be retained, yet a lower formation energy may also be achieved. In particular, the formation energy of Eu may be lowered by 0.02-0.30 eV when codoped with group IV cations, and may be lowered by 0.78-0.95 eV when codoped with group VII anions (see SI). This is due to the charge transfer from the donor's *p* states to Eu *f* states, which stabilizes the total energy of the doped system. Also, the formation energy of Eu generally can be lowered more when codoped with anions. A more significant reduction in the formation energy for the anion donors is because they have shorter distances to Eu, which is on the cation-Bi site. Thus, the anion donors can form a stronger binding with Eu due to Coulomb attraction, which is the strongest for F. After all, Eu is still slightly smaller than Bi. Hence, a compressive stress on the host can be beneficial for its incorporation. In this regard, F has the smallest atomic radius and is expected to bring in the largest local compression effect on the host to enhance Eu incorporation[33].

In summary, our study reveals several 4*f* TMs as the promising magnetic dopants for long-range stable FM orders in $Bi_2Se_3$. In startle contrast to widely-used 3*d* TM impurities, the magnetic moments of the 4*f* dopants are high, while their interstitial populations are low. The FM for these impurities is unveiled as originated from the exchange coupling between the high-lying partially-occupied/unoccupied impurity bands and the low-lying fully occupied bands. Codoping of magnetic and non-magnetic impurities was also proposed as a means to tune the Fermi level position into the band gap and to reduce the formation energy of Eu inside $Bi_2Se_3$.

This work is supported by direct grant from CUHK (4053084), ECS grant from HKRGC (ref No. 24300814) and startup funding from CUHK. SBZ was supported by the US Department of Energy (Office of Basic Energy Sciences) under Grant No. DE-SC00026 23.